\documentclass[12pt]{iopart}

\usepackage{graphicx}
\usepackage{amssymb}
\usepackage{braket}
\usepackage{setstack}
\usepackage{iopams}
\usepackage{color}

\begin{document}

\title[Towards optimal experimental tests on the reality of the quantum state]{Towards optimal experimental tests on the reality of the quantum state}

\author{George~C.~Knee}

\address{Department of Materials, University of Oxford, Oxford OX1 3PH, United Kingdom}
\address{Department of Physics, University of Warwick, Coventry CV4 7AL, United Kingdom}
\ead{gk@physics.org}
\vspace{10pt}
\begin{indented}
\item[]\today
\end{indented}

\begin{abstract}
The Barrett-Cavalcanti-Lal-Maroney (BCLM) argument stands as the most effective means of demonstrating the reality of the quantum state. Its advantages include being derived from very few assumptions, and a robustness to experimental error. {\color{black} Finding the best way to implement the argument experimentally is an open problem, however, and involves cleverly choosing sets of states and measurements. I show that techniques from convex optimization theory can be leveraged to numerically search for these sets, which then form a recipe for experiments that allow for the strongest statements about the ontology of the wavefunction to be made.} The optimization approach presented is versatile, efficient and can take account of the finite errors present in any real experiment. I find significantly improved low-cardinality sets which are guaranteed partially-optimal for a BCLM test in low Hilbert space dimension. I further show that mixed states can be more optimal than pure states.\end{abstract}

\section{Introduction}
The ontological status of the quantum state has long been a central question in foundational physics. If it is {\it ontic}, and therefore a true part of physical reality, then the many counterintuitive features of quantum theory (QT) remain opaque. The idea that it is {\it epistemic} -- that is, reducible in essence to a state of knowledge -- is an attractive proposition that promises to dissolve some of these issues~\cite{HarriganSpekkens2010}.  For instance, the instantaneous and discontinuous `collapse' of the wavefunction is arguably much more naturally thought of as a Bayesian update of knowledge than as a process governed by dynamical physical laws. There are a plethora of other canonically `quantum' phenomena that have an appealing explanation when adopting an epistemic interpretation -- see for example Ref.~\cite{Spekkens2007}. In recent years, there has been a flurry of progress towards understanding the feasibility of the so called `$\psi$-epistemic' programme~\cite{Leifer2014}, culminating in experimental tests~\cite{RingbauerDuffusBranciard2015}. Here I improve on the design of such experiments, so that tighter restrictions on possible $\psi$-epistemic theories may be determined in the laboratory.

In the ontological-models framework, where these notions are made precise, the preparation of a quantum state $|\phi\rangle$ is associated with the (generally random) selection of an ontic state $\lambda\in \Lambda$, with the appropriately normalised distribution of probability for the various states being written $\mu_\phi(\lambda)$. $\Lambda$ is the space of ontic states, and I denote the subset $\Lambda_\phi$ as the support of $\mu_\phi$ -- members of which are said to be `compatible' with the preparation. A projective measurement operator $|\psi\rangle\langle\psi|$, on the other hand, is associated with a conditional probability $\xi(\psi|\lambda)$ known as a response function. {\color{black}In the full framework, there are also stochastic maps on $\Lambda$ which represent, e.g. unitary transformations, but it will not be necessary to consider these here.} 

In order to constitute a model of quantum theory, the following condition must be met 
\begin{equation}
\int \mu_\phi(\lambda)\xi(\psi|\lambda)d\lambda=|\langle \psi | \phi\rangle|^2.
\label{q}
\end{equation}
The ontic state $\lambda$, therefore, stands for a variable sufficient to screen off the preparation $\mu$ from the measurement $\xi$. As long as (\ref{q}) is satisfied, tracing over the ontic states leaves the `correct' conditional probabilities \emph{\`a la} the Born rule.  Many such ontological models are possible, and they may be classified according to certain properties of the $\mu$ and $\xi$. Although couched in the older terminology of `hidden-variables', one of the most important such classifications is Bell's definition of locality~\cite{Bell1966}, from which he derived an experimentally testable inequality able to separate certain predictions of quantum theory from all possible {\it local} ontological-models~\cite{Bell1964}. Bell was therefore perhaps the first to show that a no-go theorem can provide positive insight into what an ontological model can be. Importantly, Bell's inequality was made robust to experimental error~\cite{ClauserHorneShimony1969}, and has now been subject to very strict tests~\cite{HensenBernienDreau2015,GiustinaVersteeghWengerowsky2015,ShalmMeyer-ScottChristensen2015}. {\color{black} Since QT could well be false, theoretical proofs which rely on quantum predictions do not establish (for example) the non-locality of nature itself. Experimental tests are therefore highly important: since they can indeed establish such features in a theory independent manner. } 

More recent work divides the ontological models into those where the wavefunction is {\it ontic} $\rightarrow\Lambda_\psi \cap \Lambda_\phi=\emptyset \quad\forall\psi\neq\phi$, from those which are {\it not-ontic}, or {\it epistemic} $\exists \psi\neq \phi: \Lambda_\psi \cap \Lambda_\phi\neq\emptyset$. Pusey, Barrett and Rudolph (PBR) proved a theorem that theoretically ruled out the subset of epistemic models where the preparation of two independent systems can be assumed to be represented by the product $\mu_1 \times \mu_2$~\cite{PuseyBarrettRudolph2012}. There are two reasons why the epistemic view cannot be immediately dispensed with, however. First, even in the idealized case, one may always instead drop the assumption of preparation independence (PI). Indeed, constructive examples of $\psi$-epistemic theories exist~\cite{KochenSpecker1967,LewisJenningsBarrett2012,AaronsonBoulandChua2013} which must therefore dodge PBR's no-go theorem in this way.  Second, the finite precision of any real experiment, including a recent one performed in an ion trap~\cite{NiggMonzSchindler2016}, mean that only a strict subclass of epistemic models subscribing to PI are ever experimentally falsified -- the `wriggle room' offered by laboratory imperfections makes it possible to retain {\it both} PI {\it and} {\color{black} an epistemic view of nature}. Subsequent theoretical studies have argued for the reality of the quantum state by imposing further assumptions on the set of ontological models~\cite{PatraPironioMassar2013,ColbeckRenner2012,Hardy2013} -- for a comprehensive review, see Ref.~\cite{Leifer2014}. 

Maroney introduced a classification of ontological models which generalises the ontic/epistemic dichotomy~\cite{Maroney2012}, so that one can begin to constrain the extent to which a model may be epistemic. Crucially, Maroney's theorem does not rely on PI, and thus its implications cannot be escaped by discarding it. Maroney's idea, later made noise tolerant by Barrett et al (BCLM)~\cite{BarrettCavalcantiLal2014}, springs from a very particular motivation for $\psi$-epistemicism: namely, the impossibility of discriminating non-orthogonal quantum states. In quantum theory, this feature depends on the quantity
\begin{equation}
\omega_Q(|\psi\rangle,|\phi\rangle):= 1-\sqrt{1-|\langle \psi | \phi\rangle|^2}\nonumber;
\end{equation}
a measure of the overlap of the quantum states in Hilbert space. It is related to the maximum probability with which the two states can be distinguished in a single shot experiment~\cite{Fuchs1996}. An analogous quantity 
\begin{equation}
\omega_C(\mu_\psi(\lambda),\mu_\phi(\lambda)):=1-\frac{1}{2}\int |\mu_\psi(\lambda)-\mu_\phi(\lambda)|d\lambda\nonumber
\end{equation}
applies to the underlying ontological model: it measures the extent to which two preparations `overlap' in the ontic statespace.
A partially-$\psi$-epistemic model is defined by the relation:
\begin{eqnarray}
k(\psi,\phi)&:= \frac{\omega_C(\mu_\psi(\lambda),\mu_\phi(\lambda))}{\omega_Q(|\psi\rangle,|\phi\rangle)},\quad \langle\psi|\phi\rangle \neq 0 \label{classification}.
\end{eqnarray}
The proportionality constant $k(\psi,\phi)$ will be the main subject of our study. If $k(\psi,\phi)=0$, there is no overlap of distributions in the ontic statespace, irrespective of the `closeness' of the quantum states they represent: the model is $\psi$-ontic. On the other hand, since it is required that  $\omega_C(\mu_\psi(\lambda),\mu_\phi(\lambda)) \leq \omega_Q(|\psi\rangle,|\phi\rangle)$~\cite{NiggMonzSchindler2016}, a model with $k(\psi,\phi)=1$,  is said to be `maximally $\psi$-epistemic'. 

The remainder of this paper is structured as follows: in Section \ref{degree}, I provide an operational interpretation for intermediate values of $k(\psi,\phi)$; in Sections \ref{bclm} and \ref{optim} I recapitulate the BCLM argument and describe the search for optimal experimental implementations as a nonlinear optimization problem. I present an algorithm for this purpose in Section \ref{algo}, along with several numerical results in Section \ref{numer}. There I show how the algorithm can tailor the design of an experiment by accepting the typical error rate as input (in Section \ref{noise}). Conclusions are drawn at the end of the paper, and {\color{black}the appendices contain details of a) an algorithmic subroutine, b) a mixed-state result that performs more optimally than any known set of pure states of the same size, and c) algorithm runtime statistics.}

\section{Degrees of epistemicness}
\label{degree}
The task of discriminating $|\phi\rangle$ from $|\psi\rangle$ in a single shot has a minimum probability of making an error (i.e. guessing the wrong state) given by 
$
P^{\mathrm{error}}_Q=\frac{\omega_Q}{2};
$
relating to the case where one makes use of the best measurement available under the constraints set by quantum theory~\cite{NielsenChuang2004}\footnote{When referring to minimum error distinction probability, we assume throughout that there is no prior information available to the state discriminator.}. A hypothetical {\it omniscient} measurement is described by a continuous set of Dirac delta functions $\xi(\lambda'|\lambda)=\delta(\lambda'-\lambda)$: it strongly violates the aforementioned constraints and reveals complete and perfect information about $\lambda$. Such a measurement, which is the `best' that is logically possible, has an error probability\begin{equation}
P^{\mathrm{error}}_\infty=\frac{\omega_C}{2}=k(\psi,\phi)\frac{\omega_Q}{2}=k(\psi,\phi)P^{\mathrm{error}}_Q.
\end{equation}
$k(\psi,\phi)$ therefore represents the improved state-discrimination error-probability of an omniscient observer, compared with the observer whom is constrained by quantum mechanics. In situations where the former would be superior, this necessitates a certain property of the response functions in the ontological model termed `deficiency' by Harrigan and Rudolph~\cite{HarriganRudolph2007}. Deficiency means a measurement of $|\psi\rangle\langle \psi|$ will respond to ontic states $\lambda$ which are not compatible with  $\mu_\psi$.  If $k(\psi,\phi)=0$, an ontological model of quantum theory must be maximally deficient -- that is to say, the inequality
\begin{eqnarray}
\int_{\Lambda \backslash \Lambda_\psi} \mu_\phi(\lambda)\xi(\psi|\lambda)d\lambda\leq  \int_{\Lambda} \mu_\phi(\lambda)\xi(\psi|\lambda)d\lambda = |\langle \psi|\phi\rangle|^2
\end{eqnarray}
will be saturated. A fuller discussion of deficiency is given in~\cite{Ballentine2014}.
\section{Bounds on $k(\psi,\phi)$}
\label{bclm}
 BCLM showed how $k(\psi,\phi)$ can be bounded without relying on PI~\cite{BarrettCavalcantiLal2014}. In Hilbert space dimension $d$, consider a reference state $\ket{c}$ along with a set of $n$ states $\mathcal{S}=\{\ket{\psi_i}\}_{i=1}^n$. Consider further a set of projective operators $\mathcal{M}=\{|ijk\rangle\langle ijk|;i=1\ldots n,j=1\ldots i-1, k=1,2,3\}$, such that associated to each pair of states $\{|\psi_i\rangle,|\psi_j\rangle\}$ in $\mathcal{S}$, there is a triple of orthogonal projectors $\{|ij1\rangle\langle ij1|,|ij2\rangle\langle ij2|,|ij3\rangle\langle ij3|\}$ in $\mathcal{M}$ that define a three-outcome measurement. We will find it useful to define $\mathcal{A}_{ij}:=|\langle ij1|\psi_i\rangle|^2+|\langle ij2|\psi_j\rangle|^2+|\langle ij3|c\rangle|^2$, which is a measure of the {\it antidistinguishability} of the triple of states $|\psi_i\rangle,|\psi_j\rangle,|c\rangle$. This concept is at the heart of many $\psi$-ontology theorems: when $\mathcal{A}_{ij}=0$ the triple is said to be perfectly antidistinguishable, and then the measurement $\{|ijk\rangle\langle ijk|\}_{k=1,2,3}$ will conclusively exclude (in a single shot) the possibility that one of the triple of states was prepared~\cite{BandyopadhJainOppenheim2014}. Caves, Fuchs and Schack have derived the necessary and sufficient conditions for perfect antidistinguishability (which they call PP-incompatibility) of a triple of states, which depend only on the three inner products between pairs in the triple~\cite{CavesFuchsSchack2002}. 

Next, define $k_0:=\min_j k(c,\psi_j)$. Then, if QT predictions are correct, the relation 
\begin{eqnarray}
k_0 \leq \frac{1+ \sum_{i>j}\mathcal{A}_{ij}}{\sum_i   \omega_Q(|c\rangle,|\psi_i\rangle)}
\end{eqnarray}
follows from (\ref{classification}) and from the Bonferonni inequality, as BCLM show~\cite{BarrettCavalcantiLal2014}. 
The remainder of this paper is concerned with finding small $\mathcal{S}$ and $\mathcal{M}$ such that if QT is approximately correct, they would lead to the lowest upper bound on $k_0$. An upper bound $k_0\leq x$ implies that there exists at least one state in $\mathcal{S}$ which has a degree of epistemicness with respect to $c$ no greater than $x$. One may optionally make additional assumptions to extrapolate this to a stronger claim: for example, Lipschitz continuity~\cite{Leifer2014} would require that $k(\psi,\phi)\leq x\quad \forall \psi,\phi$. In the course of our search, we will limit ourselves to finite values for $d$ the Hilbert space dimension and $n$ the number of states, considered here as experimental resources to be spent frugally.  

\subsection{Existing families of states}
Before presenting my results below, it is prudent to survey currently-available solutions the problem which will serve as benchmarks for our algorithmic approach. BCLM supplied for every $d\geq4$ and also power prime, a set of $n=d^2$ states, satisfying $\mathcal{A}_{ij}=0\quad\forall i,j$ which leads to the bound $k_0<2/d$~\cite{BarrettCavalcantiLal2014}. When $d\geq4$ is not power prime, their results lead to the bound $k_0<4/(d-1)$. Relaxing the need for exact antidistinguishability enabled them to find $k_0\leq0.95$ for $d=3, n=9 $. Next, Leifer showed an exponential decay in $d$ (for $d\geq3$) by using a set of $n=2^{d-1}$ Hadamard states~\cite{Leifer2014a}: this achieves the bound $k_0<4d/2^d$~\cite{Branciard2014}. Branciard provided a nonconstructive proof that $k_0\leq 8/n^{(d-3)/(d-2)}$, which displays a decay in $n$ for any $d\geq4$, as well as a number of constructive solutions~\cite{Branciard2014}. Where these previous results provide bounds on $k_0$ for low $n$ and $d$, the approach I present below was able to match or better the bound. The best bound achieved experimentally at the time of writing is  $k_0\lessapprox69\%$~\cite{RingbauerDuffusBranciard2015}.  Achieving $k_0\leq 50\%$  would seem like the next major milestone, where the classical overlap is doing less than half of the necessary work in explaining the indistinguishability of non-orthogonal quantum states.

\section{Optimization}
\label{optim}
 {\color{black} We wish to solve
\begin{eqnarray}
\underset{\mathcal{S}\times\mathcal{M}}{\mathrm{minimize}}\qquad&
\frac{1+ \sum_{i>j}\mathcal{A}_{ij}}{\sum_i   \omega_Q(|c\rangle,|\psi_i\rangle)} \nonumber \\
\mathrm{subject\,to}& \langle \psi_i|\psi_i\rangle = 1\quad i=1,\ldots n \nonumber\\
& \langle ijk|ijk\rangle = 1\quad i=1,\ldots, n\quad j=1,\ldots, i\quad k=1,2,3 \nonumber\\
& \langle ij1|ij2\rangle =\langle ij1|ij3\rangle =\langle ij2|ij3\rangle = 0\quad i=1,\ldots, n\quad j=1,\ldots, i \nonumber\\
\label{intitPROBLEM}
\end{eqnarray}
Here all optimisation variables are considered as unnormalised vectors in $\mathbb{C}^d$, with the necessary normalisation and orthogonality conditions shown explicitly above as constraints. An analytic, global solution to this problem seems intractable: but how difficult is the numerical optimization problem at hand? A naive answer is found by counting the number of real parameters necessary to describe a solution to the problem.  One can do better than simply taking the real and imaginary components of each vector, since some of the constraints imply these parameters are not independent. }For example, the hyper-spherical parameterisation of pure states requires only $d-1$ polar angles and $d-1$ phase angles~\cite{Hedemann2013} (making $2d-2$ in all), with normalisation constraints automatically satisfied. A projective measurement is defined by a $d\times d$ unitary matrix: in close analogy to the argument above concerning states, a unitary matrix may be parameterised by $d^2-1$ angles \cite{Hedemann2013}, and describes an orthonormal measurement basis by construction. Thus, even with a smart parameterisation one must solve a 
$
(2d-1)n+(d^2-1)(n^2-n)/2
$
dimensional optimization problem.

Various heuristics can inform our search. For example, to make each $|\psi_i\rangle$ as close as possible to $|c\rangle$, whilst at the same time making each state far from all of the others so that they can be (approximately) anti-distinguished. The objective function is nonlinear, nonconvex and is pocked with local optima: gradient descent will therefore most likely get `stuck' in a feasible region with sub-par performance. Brute force methods would be intractable: note that, on a coarse grid dividing each angle into only $g$ discrete values, one must evaluate the objective function $g^{36}$ times in the problem instance of $d=n=3$.  Global approaches include simulated annealing, particle swarm and associated techniques. But a more powerful approach can be implemented by capitalising on a specific structure of the problem. 

\subsection{Exploiting convexity}
\label{algo}
\begin{figure}
\begin{center}
\includegraphics[width=14cm]{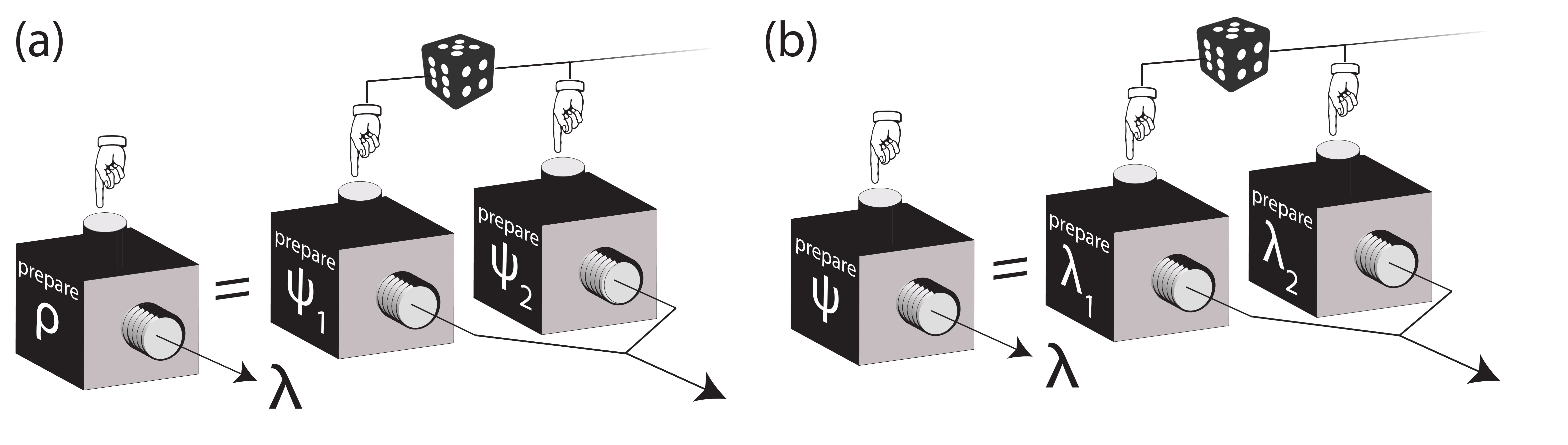}
\end{center}
\caption{In the ontological models framework, preparations may be thought of as `black boxes' that simply produce a $\lambda$ according to the distribution $\mu_\psi(\lambda)$. (a) A mixed state $\rho =\sum_i p_i |\psi_i\rangle\langle\psi_i|$ can be thought of as the net preparation when several pure preparations are wired together with a probabilistic switch. The resulting probability density over ontic states is the convex combination of the component distributions: $\mu_\rho(\lambda)=\sum_ip_i \mu_{\psi_i}(\lambda)$. (b) In much the same way, each component preparation labelled by a pure state $\psi_i$ can be thought of itself as a compound preparation, where a probabilistic switch selects from a number of deterministic preparations of $\lambda$, so that the net preparation is described by $\mu_\psi(\lambda)$. For the purposes of illustration, here $\Lambda$ is represented as a discrete space; although it is often (and in eq (1)) thought of as a continuous space.}
\label{mixing}
\end{figure}

One may `lift' the problem from a search over the set of  $n$ statevectors $\mathcal{S}$ to the set of $n$ density matrices $\mathcal{D}=\{\rho_i\}_i$ and from the set of $3(n^2-n)/2$ projective operators $\mathcal{M}$ to the set of the same number of positive operators $\mathcal{P}=\{E_{ijk}\}_{ijk}$. Let us take a moment to interpret what this might mean. A Positive Operator Valued Measure (or POVM) is a set of positive operators that sum to the identity operator. It constitutes the most general description of a quantum measurement. Likewise, a mixed state (being an Hermitian, trace-one operator) is the most general description of a quantum state. If we interpret a mixed state as a proper mixture of other states, then the corresponding preparation in the ontological model is the random selection of one of a number of preparations, each itself a random selection of $\lambda$. See Figure \ref{mixing}.

The following generalisations of the quantum overlap and the Born rule
\begin{eqnarray}
  \omega_Q(|c\rangle,|\psi_i\rangle)&\rightarrow \omega_Q(\rho_c,\rho_{\psi_i})=1-\frac{1}{2}||\rho_c-\rho_i ||_* \\
  |\langle ijk|\psi_c\rangle|^2&\rightarrow \mathrm{trace}(E_{ijk}\rho_c)
\end{eqnarray}
share an important property: {\it they conserve their meanings} as twice the minimum error discrimination probability ~\cite{Fuchs1996}, and prepare-measure probability (respectively).$||\bullet||_*$ is the nuclear norm (or sum of singular values). Derivative notions such as antidistinguishability $\mathcal{A}_{ij}$ therefore also inherit their meaning correspondingly. With the generalisations in place, we can therefore write the optimization problem as 
\begin{eqnarray}
\underset{\mathcal{D}\times\mathcal{P}}{\mathrm{minimize}}\qquad&
 \frac{1+ \sum_{i>j} \mathrm{trace}(E_{ij1}\rho_{i}+E_{ij2}\rho_{j}+E_{ij3}\rho_c + \epsilon_{ij1}+\epsilon_{ij2}+\epsilon_{ij3})}{\sum_j 1-\frac{1}{2}||\rho_c-\rho_j ||_*}  \nonumber \\
\mathrm{subject\,to}& E_{ij1} +E_{ij2} +E_{ij3} = \mathbb{I}\nonumber\\ 
& E_{ijk} \succeq 0\nonumber\\
& \rho_i \succeq 0\nonumber\\
& \mathrm{trace}(\rho_i) =1.
\label{PROBLEM}
\end{eqnarray}

Here $M\succeq0$ denotes that $M$ is a positive semidefinite matrix. $\epsilon_{ijk}$ are small error parameters to be discussed shortly -- the reader may temporarily assume these to be zero. Recall that a function $f(\mathbf{x})$ is convex iff $f(\lambda\mathbf{x}_1+(1-\lambda)\mathbf{x}_2)\leq \lambda f(\mathbf{x}_1)+(1-\lambda)f(\mathbf{x}_2)$, and that a set $\mathcal{B}$ is convex iff $\mathbf{x}_1,\mathbf{x}_2\in\mathcal{B}\rightarrow \lambda\mathbf{x}_1+(1-\lambda)\mathbf{x}_2\in\mathcal{B}$ for  $\lambda \in [0,1]$ . Call a problem {\it convex} if it involves minimising a convex function over a convex set. Such problems exhibit many convenient features: if a local optimum exists, it is also a global optimum. Efficient algorithms exist to solve convex problems. 
Although (\ref{PROBLEM}) is not a convex problem, notice that the objective function is linear (hence convex) in the set of measurements when the states are fixed. Furthemore, when the measurements are fixed, the objective is convex-concave fractional in the states: then the global optimum can be found by solving a series of parametric (convex) subproblems~\cite{EnkhbatBazarsadEnkhbayar2011}. Note further that $\mathcal{D},\mathcal{P}$ are convex sets. Thus, we proceed in a manner inspired by `biconvex' problems~\cite{GorskiPfeufferKlamroth2007} (which have a very similar structure). The approach put forward here is to begin with a feasible point in $\mathcal{D}$, and then proceed to alternately optimise over $\mathcal{P}$ and then $\mathcal{D}$ again, and so on (keeping the other set fixed) until the optimal values of the two subproblems converge. This is known as alternate convex search (ACS)~\cite{GorskiPfeufferKlamroth2007}, and whilst it does not guarantee global optimality, it does tend to provide good results which are guaranteed {\it partially optimal}: that is to say, no change in $\mathcal{D}$ or in $\mathcal{P}$ alone could provide a better solution.  In order to solve the fractional subproblem of the form min  $a/b$ (when fixing the measurements and searching for states), one may use Dinkelbach's technique~\cite{Jeflea2003}, which involves solving a series of parameterised problems min $a-\theta_k b$: {\color{black} see \ref{db}}. 

\begin{figure}[!]
\begin{center}
\includegraphics[width=14.5cm]{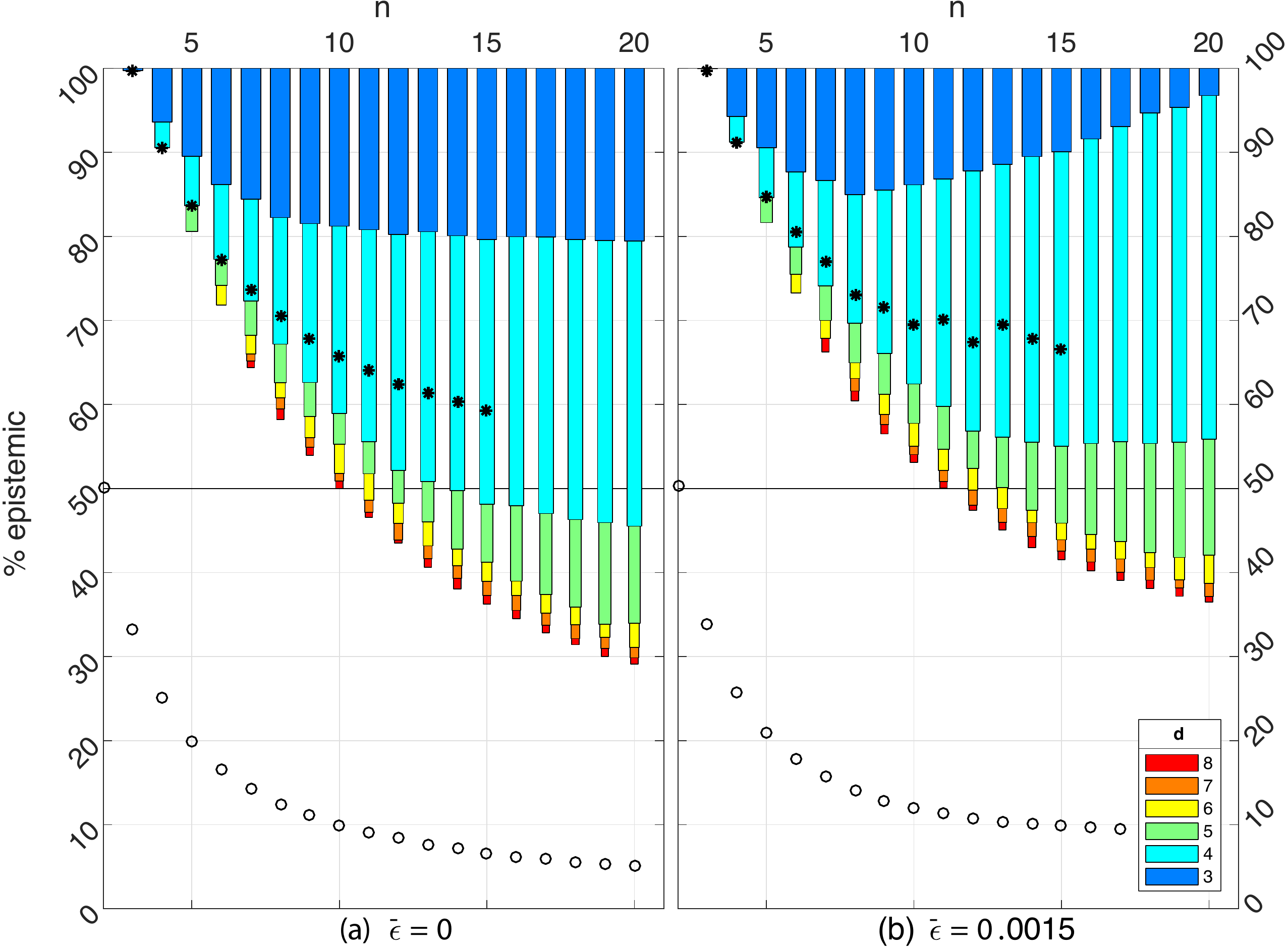}
\end{center}
\caption{\label{1pc}(a) Upper bounds on $k_0$ achievable by preparing $n+1$ quantum states each of dimension $d$, generated using an alternating search which solves convex subproblems. Increasing $d$ from $3$ to $4$ and to $5$ provide significant gains, but thereafter only diminishing returns. Open circles follow $(1+\frac{1}{2}n(n-1)3\bar{\epsilon})/n$, which is a lower bound to the optimum experiment (independent of $d$).  (b) Allowing observed statistics to deviate on average by $0.15\%$ from ideal QT predictions defines a new optimization problem, with more modest bounds still achievable. Note how an optimum $n$ emerges. Black asterisks indicate previously best-known a) theoretical and b) experimental bounds from Ringbauer et al. in $d=4$ (with $\bar{\epsilon}\sim 0.0015$)~\cite{RingbauerDuffusBranciard2015}.}
\label{fig}
\end{figure}

\section{Numerical results}
\label{numer}
There exist many software packages for solving the convex problems that arise during my algorithm. {\color{black} Because the algorithm will not always settle on the same solution if seeded with different starting points,} I ran the algorithm several times, using the {\sc cvx} package~\cite{GrantBoyd2014} for each subproblem: the best numerical results for $\epsilon_{ijk}=0$, $d=3,\ldots,8$ and $n=3,\ldots,20$ are shown in Figure \ref{fig}a. The algorithm was seeded with random pure states. Corresponding {\sc matlab} files with the states and measurements are available {\color{black} in the supplementary material}, as is the code needed to find partially optimal solutions for any $n, d$. {\color{black}The results show that an experiment showing $k_0\leq 50\%$ is possible with current technology such as linear optics~\cite{RingbauerDuffusBranciard2015} or ion traps~\cite{NiggMonzSchindler2016}. The performance of the algorithm is discussed in \ref{perf}.}

Some comments about the choice of reference state are in order. Because our objective function is invariant under a simultaneous unitary transformation of all states and measurements, the reference state may be chosen arbitrarily {\it up to the choice of eigenvalues}, which determine (for example) the purity $\mathrm{trace}(\rho_c^2)$. Because of the linearity of the subsearch over $\mathcal{P}$, the algorithm will return projective (i.e. extremal) POVMs: our generalisation to positive matrices can thus always be thought of as a purely mathematical trick, since the measurements will always correspond to a projective measurement (and it is very important that the measurement has only 3 outcomes~\cite{DuttaPawlowskiifmmode-Zelse-Zfiukowski2015}). Importantly, no such property guarantees that the search over $\mathcal{D}$ will return pure states.  The results so far correspond to a pure reference state, which caused the solution states $\rho_i$ to also all be pure. However, one can in fact include the reference state into the search space $\mathcal{D}\rightarrow \{\rho_c,\{\rho_i\}_i\}$, thereby leaving the purity as a free parameter to be optimised over. Whilst in principle this could only lead to better results, in practice it can mean the algorithm is more likely to get stuck in local optima. However, in the appendix I present ${\it mixed}$ states and measurements which lead to a $n=d=3$ bound of $k_0\leq 1.2018/1.5003 \approx 0.8011$, a marked improvement over known bounds with only pure states.

\subsection{Noise}
\label{noise}
Inspecting Figure \ref{fig}a, it seems that increasing $n$ is an easy way to improve an experiment -- but as long as the errors $\epsilon_{ijk}\neq0$ (which captures the realistic situation where the QT predictions are not precisely reflected in the experiment) they will accumulate and spoil the trend.  Let $\bar{\epsilon}=2\sum_{i>j}[ \epsilon_{ij1}+\epsilon_{ij2}+\epsilon_{ij3}]/(3n^2-3n)$ be the average error. Note that the objective function in (\ref{PROBLEM}) is monotonically increasing in $\bar{\epsilon}$ . Note also that my algorithm can adjust the tradeoff between the numerator and denominator to achieve states which are in general less optimal for the noise-free case, but more robust to error than the noise-free optima. For numerical results relating to $\bar{\epsilon}=0.15\%$, see Figure \ref{1pc}b.

Let a bound on $k_0$ be given by the ratio $A/B$, where $A$ is the (noiseless) numerator of (\ref{PROBLEM}), and $B$ the denominator. Then, the maximum tolerable error is $\bar{\epsilon}_{\mathrm{MAX}}=2(B-A)/(3n^2-3n)$. Surprisingly, the mixed state bound I found by including the reference state in the search is extremely robust to noise, tolerating $\bar{\epsilon}_{\mathrm{MAX}}\approx0.033$,  whilst the previous best $n=d=3$ (pure state) solution~\cite{Branciard2014} could tolerate $\bar{\epsilon}_{\mathrm{MAX}}=0.0006$. This constitutes, a $\sim50$-fold improvement in robustness for the same experimental resources, contradicting the widely held feeling that extremely high precision experiments are necessary to show the reality of the quantum state.

\subsection{Lower bounds} A loose lower bound on the global optimum of (\ref{PROBLEM}) (i.e. the best possible BCLM experiment) can be obtained through convex relaxation. Replace the bilinear terms in the numerator with zero: then the problem is convex and actually solvable analytically by putting $\rho_i=\rho_c\quad\forall i$. The BCLM argument therefore cannot hope to find a bound on $k$ any lower than  $(1+\frac{1}{2}n(n-1)3\bar{\epsilon})/n$. Tighter convex relaxations would be very useful if found. 

\section{Conclusions and directions}
The technique for finding BCLM experiments presented here is very flexible. Besides the results showcased above -- the improved zero-error bounds and improved finite-error bounds for low $n$ and $d$ -- there are additional applications of the method. An experimentalist, armed with an estimate of the typical precision available in her laboratory setup, can use my algorithm with this quantity as input -- and therefore extract the optimum $n$ and appropriate sets $\mathcal{D},\mathcal{P}$. 

It is at first quite surprising that non-extremal states can be a better choice in arguing for the reality of the quantum state. It is counter intuitive because i) for many quantum information processing applications, mixed states will perform strictly worse than pure states, and ii) since mixing states together introduces `artificial' ignorance into the problem one would then expect the states to become more epistemic. But since we are interested in the ratio of quantum to classical overlaps, we are not interested in the absolute measure of epistemicness, but rather the upper limit on how close the epistemicness of the ontological model can come to explaining the overlap at the quantum, or operational level. Intuitively, increasing the mixedness of states can make them less distinguishable: increasing their overlap $\omega_Q$ and therefore decreasing our objective function. Fundamentally, a mixed state $\rho$ in quantum mechanics will still have to correspond to some distribution $\mu_\rho$ in an ontological model (see Figure~\ref{mixing}). Such preparations also exhibit the property that (except in special cases) one cannot distinguish them with a single shot measurement. The epistemic interpretation is therefore just as compelling for mixed states as for pure states. One difference is that even within quantum mechanics there is an `ignorance' interpretation of a mixed state as a (non-unique) convex combination of pure states.  But since our goal is to show that there is an ever widening explanatory gap that the epistemic interpretation fails to breach, this is by no means a get out clause to the argument. Needless to say that the often-cited belief that `there are no pure states in the laboratory' should cause $\rho$-epistemicism to be taken just as seriously, nay, {\it more seriously} than $\psi$-epistemicism.

{\color{black} Each of the mathematical elements of an ontological model (ie. the preparations, transformations and measurements) should in-principle carry labels that allow for contextuality~\cite{Spekkens2005,LeiferMaroney2013}. This is because multiple, physically distinct preparations (transformations, or measurements) may be identified in QT, but one is generally not warranted in identifying their representation in the ontological model. Doing so amounts to making an additional (and spurious~\cite{Spekkens2005,LeiferMaroney2013}) assumption of non-contextuality.  It is therefore paramount in any experimental test relating to ontological models (such as those proposed by PBR and BCLM) that the very same physical procedure be used at any point where a definite preparation or measurement is repeatedly called for. The temptation to use distinct procedures which would be equivalent in QT is perhaps greater for objects with multiple convex decompositions, such as mixed states: but it should be resisted all the same, because it would introduce a contextuality loophole.}

Further, it is simple to apply additional constraints which will not spoil the properties of (\ref{PROBLEM}). For example;  the expression $||\rho_c-\rho_j||_*$ can be upper bounded, or $\mathrm{trace}(\rho_c\rho_j)$ can be bounded from above (and/or below if choosing $\rho_c$ to be fixed, and not part of $\mathcal{D}$). These constraints might help provide further theoretical insight into the nature of ontological models for QT.  Similar constraints might help experimentalists search for those certain states and measurements that are easy to prepare with high fidelity. It is also possible to seed the algorithm a more structured and informed initial feasible point from which to search from -- rather than the random feasible points I chose. My algorithm will return a solution at least as good as its input, and can therefore be used to `polish' any solution found by other means.

It is possible in future that deterministic global optimization techniques~\cite{Floudas2013} can be applied to this problem, and provide a certificate of \emph{global} optimality (rather than just partial optimality) for $\mathcal{D}$ and $\mathcal{P}$. Such a certificate would be a akin to a ``Tsirelson's bound''~\cite{Cirelson1980} -- and would provide some much-desired certainty in the search for optimal BCLM experiments.

\ack
I would like to thank Owen Maroney and Andrew Briggs for helpful discussions, and Ronnie Hermens and Cyril Branciard for feedback on an earlier version of this manuscript. This publication was made possible through the support of a grant from Templeton Religion Trust. The opinions expressed in this publication are those of the author and do not necessarily reflect the views of Templeton Religion Trust. This work was  supported by the Royal Commission for the Exhibition of 1851.

\appendix
{\color{black}
\section{Dinkelbach's technique}
\label{db}
Dinkelbach's technique is an iterative method of solving a convex-concave fractional problem: here I follow Jeflea's description~\cite{Jeflea2003} -- the interested reader should consult this reference for more details. Consider the problem
\begin{eqnarray}
{\mathrm{minimize}}\qquad&
 \frac{a(x)}{b(x)}: x\in X ,
 \label{dbf}
\end{eqnarray}
where $a(x)$ and $b(x)$ are (respectively) convex and positive, concave functions and $X$ is a compact convex set.
%Jagannathan's theorem (1966) sates that $y\in X$ is an optimal solution for the above problem iff $y$ is an optimal solution for 
%\begin{eqnarray}
%\underset{X}{\mathrm{minimize}}\qquad&
% a(x)-\frac{e(y)}{a(y)}b(x).
%\end{eqnarray}
Dinkelbach developed the following algorithm which he proved to converge on the globally optimal solution to this problem~\cite{Dinkelbach1967}:
\begin{enumerate}
\item Initialise $x_1\in X$. Let $k=1$. \\
\item Let $\theta_{k}=a(x_k)/b(x_k)$. Find the solution $x_{k+1}$ to  $\min_X \{a(x)-\theta_k b(x)$\}. \\
\item If  $a(x_{k+1})-\theta_k b(x_{k+1}) = 0$, stop and $x_{k}$ is optimal. Otherwise  $k=k+1$ and go to previous step. 
\end{enumerate}
The subproblem of my alternating search approach that involves fixing the measurements and searching for states is of the form (\ref{dbf}): it may therefore be solved with Dinkelbach's method. 
}

\section{Mixed states solution}
 With pure states in the $n=d=3$ case, my algorithm never returned a bound better than $k_o\leq 0.9964\ldots$ {after many trials}. One therefore has the suspicion that this is in fact the global optimum of the problem when the reference state is set to a pure state. Inserting the reference state into the search space removes unnecessary constraints, for example that it must have a certain purity. The algorithm found:

\begin{eqnarray}
\rho_c&\phantom{-} = 
\left(
\begin{array}{lll}
\phantom{-}0.36238&-0.05820-0.19604i&\phantom{-}0.04384-0.07843i\\
-0.05820+0.19604i&\phantom{-}0.19657&-0.09383-0.09519i\\
\phantom{-}0.04384+0.07843i&-0.09383+0.09519i&\phantom{-}0.44104\
\end{array}
\right)\nonumber\\\nonumber\\
\rho_1&\phantom{-} = 
\left(
\begin{array}{lll}
\phantom{-}0.46135&\phantom{-}0.05756-0.01170i&-0.08267-0.03289i\\
\phantom{-}0.05756+0.01170i&\phantom{-}0.43220&\phantom{-}0.17263+0.06280i\\
-0.08267+0.03289i&\phantom{-}0.17263-0.06280i&\phantom{-}0.10645
\end{array}
\right)\nonumber\\\nonumber\\
\rho_2&\phantom{-} = 
\left(
\begin{array}{lll}
\phantom{-}0.31986&-0.04357+0.08822i&\phantom{-}0.07957+0.21626i\\
-0.04357-0.08822i&\phantom{-}0.42252&-0.07043+0.09595i\\
\phantom{-}0.07957-0.21626i&-0.07043-0.09595i&\phantom{-}0.25762
\end{array}
\right)\nonumber\\\nonumber\\
\rho_3&\phantom{-} = 
\left(
\begin{array}{lll}
\phantom{-}0.16537&\phantom{-}0.09443+0.20002i&-0.04010-0.04801i\\
\phantom{-}0.09443-0.20002i&\phantom{-}0.34712&\phantom{-}0.06583-0.02614i\\
-0.04010+0.04801i&\phantom{-}0.06583+0.02614i&\phantom{-}0.48752
\end{array}
\right)\nonumber\\\nonumber\\
U_{12}&\phantom{-}= 
\left(
\begin{array}{lll}
-0.21308-0.34918i&-0.73262-0.03220i&\phantom{-}0.28060-0.46494i\\
\phantom{-}0.41066+0.24840i&-0.04936-0.42176i&-0.50882-0.57485i\\
-0.72704-0.27156i&\phantom{-}0.23588-0.47568i&-0.34016-0.00688i
\end{array}
\right)\nonumber\\\nonumber\\
U_{13} &\phantom{-}= 
\left(
\begin{array}{lll}
-0.09331-0.10812i&-0.82454-0.21487i&\phantom{-}0.20553-0.45970i\\
-0.23642+0.21660i&\phantom{-}0.25851-0.45038i&-0.59552-0.52237i\\
\phantom{-}0.17509-0.91986i&\phantom{-}0.03121-0.05766i&-0.34483-0.00239i
\end{array}
\right)\nonumber\\\nonumber\\
U_{23} &\phantom{-}= 
\left(
\begin{array}{lll}
-0.45263+0.11154i&-0.69730+0.14663i&\phantom{-}0.45256-0.26488i\\
-0.27154-0.06431i&\phantom{-}0.21650-0.52185i&-0.19325-0.75206i\\
\phantom{-}0.02490-0.83917i&-0.20971+0.35931i&-0.30627-0.16823i
\end{array}
\right)\nonumber%\\
%\phantom{x}
\end{eqnarray}
which gives $k_0\leq 1.2018/1.5003 \approx 0.8011$. Note that here a measurement is represented by a unitary matrix $U_{ij}$, where each column is a pure state $|ijk\rangle$ corresponding to the  projector $|ijk\rangle\langle ijk|=E_{ijk}$. % Previously, with pure states we had the bound $k_0\leq 1.5051/1.5106 \approx 0.9964$.

{\color{black}
\section{Performance of the algorithm}
\label{perf}
To gain an idea of the runtime of the algorithm, and the topology of the optimisation landscape, observe Figure \ref{profile}. For the purposes of making this figure, the algorithm was run in $d=3$, a total of 1000 times for each value of $n=3,4,5,6,7$ on  a 3.6 GHz Intel Core i7-4790 running openSUSE linux, Matlab R2016a and {\sc cvx}. The results suggest that the mean runtime scales linearly with $n$, although generally the algorithm will find one of many distinct local optima depending on the initial random seed.
 \begin{figure}[!]
\begin{center}
\includegraphics[width=14.5cm]{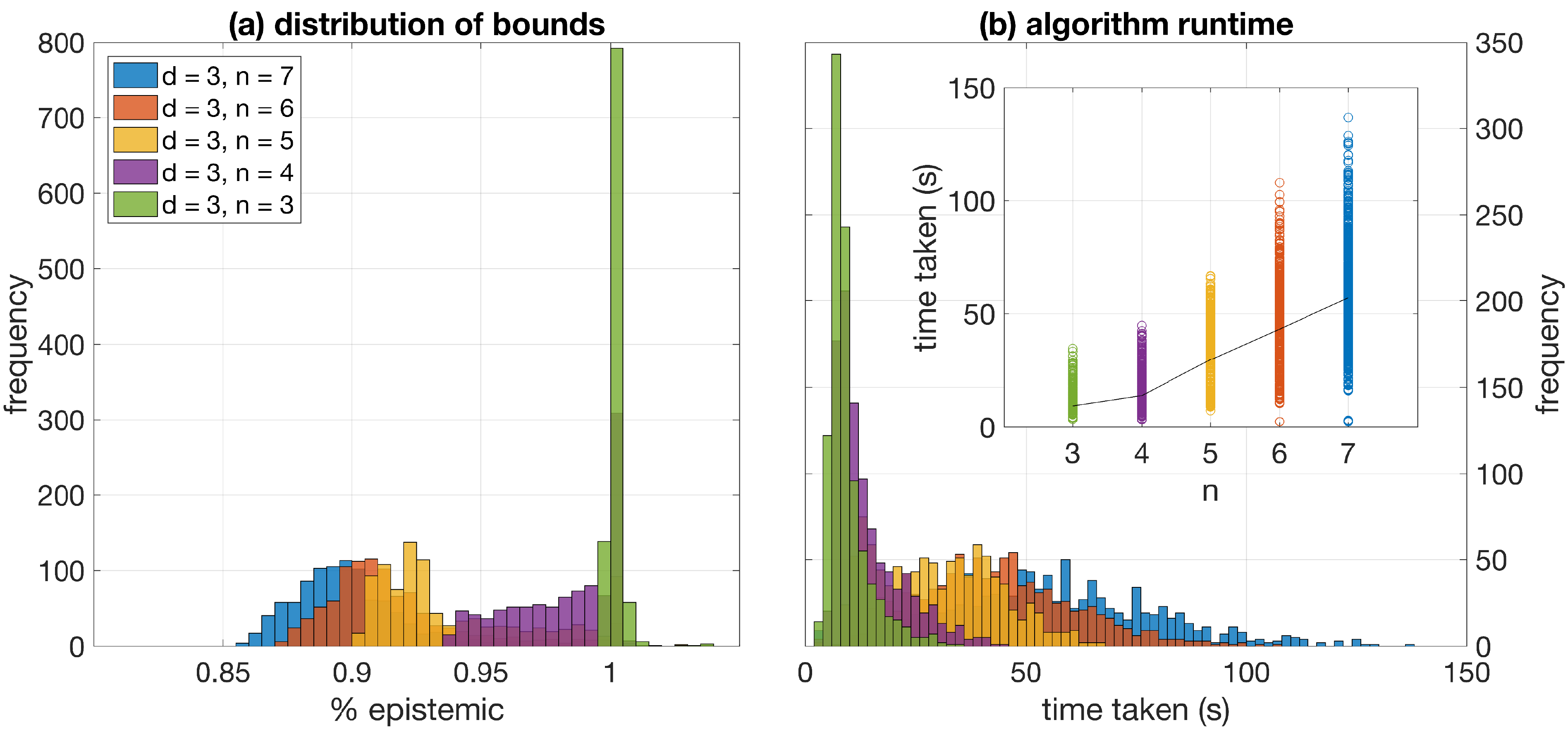}
\caption{\label{profile} Histograms of the distribution of (a) output objective function value and (b) running time for 1000 trials on a single CPU implementation of the algorithm. When the algorithm is seeded randomly, it takes a variable amount of time to finish, and settles on distinct (generally non-globally optimal) solutions. The inset of (b) shows the same data with the black line tracing the mean value, which is linear for $d\geq4$. }
\end{center}
\end{figure}
}
\section*{References}

%  BAKED IN 
%\bibliographystyle{h-physrev}
%\bibliography{/Users/georgeknee/Documents/paper_library/gck_full_bibliography}

\end{document}